\documentstyle[11pt]{article}

\def\rr{{\cal R}}
\def\p{\partial}
\def\f{\frac}
\def\om{\omega}

\newcommand{\be}{\begin{equation}}
\newcommand{\ee}{\end{equation}}

\begin{document}
\begin{titlepage}

\vspace*{1cm}

\begin{center}
{\LARGE\bf Quantum radiation from spherical mirrors} \\
\vskip 10mm

{\large L.Hadasz${}^a$, M.\,Sadzikowski${}^b$ and  
P.\,W\c{e}grzyn${}^a$}$^{\star}$ \\
\vskip 3mm
${}^a$Institute of Physics, 
Jagiellonian University Reymonta 4, PL-30\,059 Krak\'ow, Poland,\\
${}^b$Institute of Nuclear Physics,
Radzikowskiego 152, PL-31\,342 Krak\'ow, Poland.
\end{center}

\vskip 10mm

\begin{center}

\begin{abstract}

We consider mirrors of the spherical shape, that can expand or contract.
Due to the excitation of the vacuum around, some
spherical waves radiated from vibrating mirrors are encountered.
Using experience from well-known literature on studies of two-dimensional conformal models,
we adopt a similar framework to investigate such quantum phenomena in four dimensions. 
We calculate quantum averages of the energy-momentum tensor for s-wave approximation.
\end{abstract}
\end{center}

\vspace{\fill}

\noindent
\begin{tabular}{l}
TP-JU 14/97 \\
November 1997 \\
hep-th/9803032
\end{tabular}

\vspace*{2cm}
\noindent
\underline{\hspace*{10cm}}

\noindent
$^{\star}$ E-mail: wegrzyn@thrisc.if.uj.edu.pl

\end{titlepage}

\section{Introduction}

According to Quantum Field Theory, a quantum vacuum state is the lowest
eigenstate of the energy operator. In the language of the second quantization,
it refers to the state with no particles. Sometimes, this simple basis state 
can become a quite complicated mathematical object if we try to describe it
using observables related to quanta of elementary free fields.
The fundamental observation is that all quantized fields
have zero-point (vacuum) energies of size $h\nu/2$, where eigenfrequencies
can be subject to some complicated equations defining the actual physical system.
In other words, the physical vacuum can yield a complicated state
involving virtual photons, electrons and other elementary quanta.
Vacuum field  fluctuations lead to significant  physical
effects, including charge and mass renormalization, Lamb shifts, Casimir
effects or Unruh-Davies effects. The importance of vacuum fluctuations
is recognized immediately whenever some non-trivial boundary conditions
are assumed in field theoretical models.

Among many interesting problems of this kind, some theoretical models
are investigated where either a boundary of physical space or an interface
between two different media changes its shape with time. A prototype example
is a moving mirror (Unruh-Davies effect \cite{unruh}).
Induced by the motion of such objects, quanta of
massless fields are created around by vacuum fluctuations and
they can be radiated away. If a moving interface is neutral (as it
usually occurs in physical applications), such radiation is weak.
In practice, to get a significant radiation flux it is necessary
to assume  that interfaces or mirrors move with almost light-like velocities
or undergo sudden or large accelerations. These theoretical ideas
have been used to model the effects of strong gravitation on QFT or to derive
Hawking radiation \cite{birrel,brout}, to describe the squeezing effects
in quantum optics \cite{jaerey}, or to exploit the Schwinger suggestion how
to explain the phenomenon of sonoluminescence \cite{schwinger}.

In this paper, we study the (3+1)-dimensional problem of a single spherically
symmetric mirror. For the sake of simplicity, the case of massless
scalar field is considered. The mirror acts as a perfectly reflecting
infinite potential barrier, so this implies Dirichlet boundary conditions
for quantum fields. Its radius depends on time, the mirror can expand
or contract. We find the radiation flux.
Our analysis is based on standard papers devoted to the two-dimensional
moving mirrors \cite{moore,fuldav,davful}, see also recent
papers \cite{jaerey,frolov,castagnino,walker,carlitz,barton,barton2}. 
Maybe the most interesting physical situation
involves a case when a mirror oscillates with the period related to
the eigenmodes of the two-dimensional cavity. Under this resonance condition we get
that the Casimir force is enhanced \cite{law,dodonov}.
The main feature of two-dimensional
situation is that the conformal symmetry allows one to transform the problem
of time-dependent boundaries to the problem with static boundaries.
It is not possible in the four-dimensional case.
The quantum radiation of scalar particles by a moving plane mirror in the
four dimensions has been considered in \cite{ford} (for non-relativistic motions).
 This paper addresses itself
to the calculation of vacuum energy-momentum tensor in the presence
of a moving spherical mirror  in four dimensions. However, we restrict
ourselves to take account only spherical waves and spherically symmetric
quantum excitations.

\section{Spherical mirror}

The model, we are dealing with in this paper, is the quantum theory of a
massless real scalar field $\phi$ in (3+1)-dimensional Minkowski space-time.
The field is almost free, no self-interactions are present except of interactions 
with a perfectly reflecting, spherical mirror.
By a perfect mirror we mean a surface $\Sigma$ on which the field
is subjected to the Dirichlet boundary conditions,
$$
\phi|_{{}_\Sigma} = 0.
$$
Center of the mirror rests in the chosen inertial reference frame
while its radius $R$ changes in time. 

As we want to compute the energy flux carried by the particles
produced due to the motion of the mirror,
it is crucial to have a well defined notions of the vacuum state 
and of the particle both in the past and in the future.
It is possible if the mirror is initially and finally static,
so we assume that its radius changes in time according to the 
formula
\begin{equation}
R = \left\{ \begin{array}{lll}
R_0 & \mbox{for} & t \leq 0 \\
R(t) & & 0\leq t\leq t_1 \\
R_1 &  & t \geq t_1 \end{array}
\right.
\label{r2}
\end{equation}

Due to spherical symmetry of the mirror, we are able to
solve the d'Alambert equation,
\begin{equation}
\left( -\p^2_t + \Delta \right) 
\phi(t,\vec r) = 0,
\label{r1}
\end{equation}
using the method of separation of the variables in the spherical
coordinates. We get
\be
\label{l2}
\phi(t,\vec r) = \sum_{l=0}^{\infty}\sum_{m=-l}^{l}\;
          C_{lm}\phi_l(t,r)Y_{lm}(\theta,\varphi) + {\rm c.c.} \ ,
\ee
where $Y_{lm}(\theta,\varphi)$ are spherical harmonics, 
$C_{lm}$ are arbitrary complex numbers and the 
functions $\phi_l(t,r)$ are solutions of the equations
\begin{equation}
\label{l3}
\left(\f{\p^2}{\p t^2} - \f{1}{r}\f{\p^2}{\p r^2}\,r + 
\f{l(l+1)}{r^2}\right)\phi_l(t,r) = 0, \hskip 10mm
\phi_l(t,R(t)) = 0,
\end{equation}
with $l$ being a positive integer.

In the present paper we shall consider only the functions 
$\phi(t,\vec r) \equiv \phi(t,r)$ which are independent of the 
angular variables. In other words, in the partial wave expansion
(\ref{l2}) we confine ourselves to the $s$--wave contribution
with $l=0$.
This allows us to keep the technical details of the 
presented calculation at the relatively simple level. We plan
to consider the more complicated case of the higher 
partial waves in the separate article.

A general classical solution of the d'Alambert equation, assumed to be
independent of the angles 
and to vanish on the surface of the mirror, can be written in the
form
\be
\label{l4}
\phi(t,r) = \int_{-\infty}^\infty \!\!d\om\;\beta(\om){\rm e}^{-i\om t}
            \psi_\om(t,r) + {\rm c.c},
\ee

where
\be
\psi_\om(t,r) = 
\frac{1}{2ir}\left\{ {\rm e}^{i\omega (\Delta r -
\delta_R)} - {\rm e}^{-i\omega \Delta r} \right\},
\ee
and $\beta(\om)$ is an arbitrary complex function.
$\Delta r$ and $\delta_R$ are defined by the formulae
\begin{equation}
\Delta r = r - R_0, \hskip 10mm
\delta_R = 2 \left[ R(t_R) - R_0 \right].
\label{r7}
\end{equation}
$R(t_R)$ is a position of the mirror in a retarded time $t_R$, i.e.
at the past moment when the wave incoming at the present time $t$
and position $r$ was reflected from the mirror,
\begin{equation}
r - R(t_R) = t - t_R \ .
\label{r8}
\end{equation}
From the eq. (\ref{r8}) it follows that $t_R$ and $R(t_R)$ are functions of 
$t-r$.

For $t \leq 0$ the mirror is at rest and we can perform the standard
quantization procedure by imposing on the field $\hat\phi_{in}(t,r)$
(now promoted to the level of an operator) and its conjugated momentum, 
$$
\hat\pi_{in}(t,r) = \p_t\hat\phi_{in}(t,r),
$$
the canonical, equal time commutation relations written in the spherical
coordinates,
\be
\label{comrel}
\left[\hat\pi_{in}(t,r),\hat\phi_{in}(t,r)\right] = 
-\f{i}{r^2}\delta(r-r').
\ee
Subscript ``{\em in}'' serves to recall that the relations above are
valid only for $t\leq0.$ 

The field operator $\hat\phi_{in}(t,r)$
can be expanded in the basis of classical solutions of the d'Alambert
equation,
\be
\label{r10}
\hat\phi_{in}(t,r) = 
  \f{1}{2\pi}\int_{0}^\infty \!\!\f{d\om}{\sqrt{\om}}\;
  \left[b_\om{\rm e}^{-i\om t} + b^{\dag}_\om{\rm e}^{i\om t}\right]
            \psi^{(0)}_\om(r),
\ee
where
$$
\psi^{(0)}_\om(r) = \frac{\sin{(\omega \Delta r)}}{r} =
\psi_\om(t,r) \hskip 5mm {\rm for}\;\; t \leq0.
$$

The operators $b_\om, b^{\dag}_\om$ satisfy the standard commutation relations
(which follow from (\ref{comrel})),
\begin{eqnarray}
\label{r11}
\left[b_{\omega}, b_{\omega'}^{\dagger}\right] & = & 
\delta (\omega - \omega'),
\nonumber \\
&& \\
\left[b_{\omega}, b_{\omega'}\right] & = &
\left[b_{\omega}^{\dagger}, b_{\omega'}^{\dagger}\right] = 0,
\nonumber
\end{eqnarray}
The above relation  allow us to interprete corresponding operators as an
annihilation and creation operators for the field quanta. 
Consequently, we can define the state $|0\rangle_{in}$ which contains no 
particles through the equation
\begin{equation}
b_{\omega} |0\rangle_{in} = 0 \hskip 5mm \forall \; \om.
\label{r13}
\end{equation}

From the moment $t=0$, the mirror starts to move and finally, at $t=t_1,$
it reaches the shape of a sphere with the radius $R=R_1.$
In the period $0 \leq t \leq t_1$
the evolution of the field operator differs from the 
free one in (\ref{r10}) due to destorsion of the modes caused by the
interaction with the moving boundary,
\begin{equation}
\hat{\phi}(t,r) =
\frac{1}{2\pi} \int_{0}^{\infty}
\frac{d\omega}{\sqrt{\omega}} \, \left[ b_{\omega} e^{-i\omega t}
\psi_{\omega}(t,r) +
b^{\dagger}_{\omega} e^{i\omega t} \psi_{\omega}^{\ast}(t,r)
\right].
\label{r14}
\end{equation}
While the system remains in the state $|0\rangle_{in},$ it can no longer
be viewed as containing no particles. This is due to the fact that
for a generic motion of the mirror the functions $\psi_\om(t,r)$
contain for $t \geq 0$  an admixture of the negative frequency
modes $\sim\; {\rm e}^{+i\nu t},\; \nu > 0,$ and consequently
$b_\om$ cannot be interpreted as some annihilation operators for
the ``{\em out}'' vacuum.

In order to calculate the amount of produced energy and momentum
we shall compute the expectation value  of the energy--momentum
tensor $T_{\mu\nu}.$  In the considered spherically symmetric
case it has only two non-vanishing components: the energy density,
\begin{equation}
T_{tt} =
\frac{1}{2} \left\{\left( \frac{\partial \phi}{\partial t} \right)^2 +
\left( \frac{\partial \phi}{\partial r} \right)^2\right\},
\label{r15}
\end{equation}
and the radial flux of the energy (or the momentum density),
\begin{equation}
T_{tr} = \f12\left\{
 \frac{\partial \phi}{\partial t}\frac{\partial \phi}{\partial r} +
 \frac{\partial \phi}{\partial r}\frac{\partial \phi}{\partial t}
 \right\} \ .
\label{r16}
\end{equation}
The corresponding continuity equation takes the form,
\begin{equation}
\partial_t T_{tt} - \frac{1}{r} \partial_{r} T_{tr} = 0.
\label{r17}
\end{equation}

Our main aim is to calculate quantum average values of the energy-momentum tensor.
To establish a correct ultraviolet behaviour, we make use of point-splitting
regularization method ($\epsilon \rightarrow 0$),
\begin{eqnarray}
\langle T_{tt}\rangle^\epsilon & \stackrel{\rm def}{=} & 
\frac{1}{2} \ _{in}\langle0|\partial_t \phi(t,r) \partial_t \phi(t+i\epsilon,r)
+ \partial_r \phi(t,r) \partial_r \phi(t+i\epsilon,r)|0\rangle_{in}, 
\nonumber \\
&& \\
\langle T_{tr}\rangle^\epsilon& \stackrel{\rm def}{=} & 
\frac{1}{2} \ _{in}\langle0|\partial_t \phi(t,r) \partial_r \phi(t+i\epsilon,r)
+ \partial_r \phi(t,r) \partial_t \phi(t+i\epsilon,r)|0\rangle_{in}.
\nonumber
\end{eqnarray}

After straightforward calculations we obtain the following result
for the energy density:
\begin{eqnarray}
\label{endens1}
\langle T_{tt}\rangle^\epsilon & = & 
\frac{1}{32 \pi^2 r^2} \left(
\frac{4}{\epsilon^2}
- \frac{1}{3} \left\{ u+2\rr(u), u \right\}_S -
\frac{2}{r} \left\{ \frac{4(1+\dot\rr(u))[r-\rr(u)]}{\epsilon^2
+4[r-\rr(u)]^2} + \frac{\ddot\rr(u)}{1+2\dot\rr(u)}  \right\} +
\right.
\nonumber \\
&& \\
&&
\left.
+ \frac{1}{r^2}  \left\{ \ln{\left[ 1 +
\frac{4(r-\rr(u))^2}{\epsilon^2} \right]} - \ln{(1+2\dot\rr(u))}
\right\}  \right) \; + \; {\cal O}(\epsilon),
\nonumber
\end{eqnarray}
and for the radial momentum density:
\begin{equation}
\label{r22}
\langle T_{tr}\rangle^\epsilon =
\frac{1}{32 \pi^2 r^2}\left(
- \frac{1}{3} \left\{u + 2\rr(u), u \right\}_S 
- \frac{2}{r} \left[ \frac{\dot\rr(u)}{
r-\rr(u)} + \frac{\ddot\rr(u)}{1+2\dot\rr(u)}  \right]
 \right)\; + \; {\cal O}(\epsilon),
\end{equation}
where we have used an abbreviation $\rr(u) = R(t_R(u)),\; u =t-r.$
The Schwartz derivative which appears in (\ref{endens1},\ref{r22})
is defined as
$$
\left\{u + 2\rr(u),u\right\}_S \stackrel{\rm def}{=}
\f{2\stackrel{...}{\rr}(u)}{1+2\dot\rr(u)} - 
\f32\left(\f{2\ddot\rr(u)}{1+2\dot\rr(u)}\right)^2.
$$

As the formula (\ref{endens1}) contains products of
coinciding field operators taken in the same (for $\epsilon \to 0$) 
space-time point, it diverges in the limit of vanishing 
regulator. The appearance of this
divergence reflects well-known ultraviolet problems in continuum quantum field
theories. However, we should keep in mind here that the difference of energy densities
between two physically realizable situations has 
a definite meaning and thus should be finite.

Following this outline, we subtract from (\ref{endens1}) the 
(regulated with the same prescription) vacuum
energy density of the spherical wall with constant radius $R_1,$
\begin{equation}
\langle T_{tt} \rangle^\epsilon_{R_1} = 
\frac{1}{32 \pi^2 r^2} \left(\f{4}{\epsilon^2} +
\frac{1}{r^2}  \ln\left[ 1 +
\frac{4(r-R_1)^2}{\epsilon^2}\right]
- \frac{2}{r}\ \frac{4(r-R_1)}{\epsilon^2 + 4(r-R_1)^2} \right),
\label{17}
\end{equation}
and define 
\begin{eqnarray}
\label{r21}
\langle T_{tt}\rangle_{\rm ren} & = &
\lim_{\epsilon\to0}\Big(\langle T_{tt}\rangle^\epsilon - 
\langle T_{tt} \rangle^\epsilon_{R_1}\Big)
 = 
\frac{1}{32 \pi^2 r^2}\left(
- \frac{1}{3} \left\{ u+2 \rr(u), u \right\}_S  - \right.\nonumber \\
&& \\
&-& \left. 
\frac{2}{r} \left\{\frac{\dot\rr(u)\big(r-R_1\big) + \rr(u) - R_1}
{(r-R_1)(r-\rr(u))} + \frac{\ddot\rr(u)}{1+2\dot\rr(u)}  \right\}
+ \frac{1}{r^2} \ln\left[
\frac{r-\rr(u)}{(r-R_1)(1+2\dot\rr(u))} \right] \right).
\nonumber
\end{eqnarray}
The expression (\ref{r22}), which contains only products of distinct operators,
is finite in the vanishing regulator limit, thus there is no need for
any subtraction procedure,
\be
\label{l5}
\langle T_{tr}\rangle_{\rm ren}  = 
\lim_{\epsilon\to0}\langle T_{tr}\rangle^\epsilon.
\ee

The renormalized quantities (\ref{r21},\ref{l5})
satisfy the same continuity equation as their classical counterparts
(\ref{r15},\ref{r16}),
\be
\label{l6}
\p_t\langle T_{tt}\rangle_{\rm ren} -
\f{1}{r}\p_r\langle T_{tr}\rangle_{\rm ren} = 0.
\ee
It is worth to stress here that there is no contribution to the total energy
from the subtracted expression (\ref{17}).
The quadratic term produces an infinite constant, while the other
two terms in (\ref{17}) return zero when integrated out over the whole space
outside the sphere.

Using (\ref{r21}) we are finally at the position to calculate 
(for $t \geq t_1$) the total energy. We obtain surprisingly compact formula,
\begin{equation}
\label{l8}
E = \int_{t= {\rm const}}
\hspace*{-5mm}dV \; \langle T_{tt}\rangle_{\rm ren} =
\frac{1}{12\pi} \int_{R_1}^{\infty}\!\!\!dr \;
\left( \frac{\ddot\rr(u)}{1+2\dot\rr(u)}
\right)^2 \ .
\end{equation}
Let us note that respective boundary terms vanish here due to the assumption
of static mirror asymptotical states and the symmetry of the considered physical system.
We observe that the only term in Eq.(\ref{r21}) which contributes
to the total energy is just the one containing the Schwartz derivative.

In the remainder of this paper, we consider
an example being an explicitly solvable problem.
It refers to the following time-dependence of
the radius of the spherical mirror,
\begin{equation}
R(t) = \left\{ \begin{array}{lll}
R_0 & \mbox{for} & t\leq 0 \\
t + R_0 - a + a \exp{(-t/a)} & & 0 \leq t \leq \bar t \\
t + R_0 - b - t_1 + b \exp{((t_1-t)/b))} & & \bar t \leq t \leq t_1 \\
R_1 &  & t \geq t_1 \end{array}
\right.
\end{equation}
where
\begin{equation}
t_1 = \frac{(R_1-R_0)\bar t}{\bar t- a v_{max}} \ \ \ ,
\ \ \ b = - \frac{a(t_1 - \bar t)}{\bar t}.
\end{equation}
Here $a,\bar t,$ the initial and final
radii $R_0$ and $R_1$ are the parameters defining 
the ${\cal C}^1$--class function 
$R(t)$ and $v_{max}$ is a maximal velocity of the mirror
\begin{equation}
v_{max}= 1- \exp{(-\bar t/a)}.
\end{equation}

The retarded time can be calculated by solving Eq.(\ref{r8}),
\begin{equation}
t_R(u) = \left\{ \begin{array}{lll}
u + R_0 & \mbox{for} & u \leq - R_0 \\
- a \ln{[(a-u-R_0)/a]} & & - R_0 \leq u \leq a v_{max} - R_0 \\
t_1 - b \ln{[(b+t_1-u-R_1)/b]} & & a v_{max} - R_0 \leq u \leq
t_1 - R_1 \\
u + R_1 & & u > t_1 - R_1 \end{array}
\right.
\end{equation}
The basic function $\rr(u)$ is obtained immediately,
\begin{equation}
\rr(u) = t_R(u) - u,
\end{equation}
and allows to compute explicitly the 
total energy (\ref{l8}):
\begin{equation}
E = \frac{1}{24\pi} \frac{R_1-R_0}{a(R_1-R_0+a v_{max} -\bar t)}
\left[ \frac{v_{max}}{1-v_{max}^{2}} + {\rm artanh}\;v_{max} \right] \ .
\end{equation}

\section{Conclusions}

In this paper, the radiation emitted by the spherical mirror
has been considered.
We assumed the movement of the mirror lasts a finite period of time.
This assumption helps to define and interprete asymptotic spaces of physical
states uniquely. As the initial state it is considered the vacuum state
(no particles present). The vacuum is perturbed by the moving mirror,
and some flux of particles (radiation) is seen by an observer in the
laboratory frame. The radiation is a pure quantum effect.
We have calculated the vacuum expectation value of the energy-momentum tensor.
To define finite results, we adopted the time-splitting regularization
technique.
To make calculations simpler, we restrict ourselves only to spherical
waves and spherically symmetric quantum excitations.
We obtained that the radiation flux depends only upon values
of the mirror radius and its time derivatives evaluated along the
intersection of the world history of the mirror with the
observer's past light cone.

\section{Acknowledgements}

This work was supported in part by KBN grant No. 2 P03B 095 13.

\end{document}